\documentclass{article}
\usepackage{setspace}
\usepackage{amsthm}
\usepackage{rotating}
\theoremstyle{definition}
\newtheorem{definition}{Definition}[section]
\usepackage{longtable}
\usepackage{url}
\usepackage{algorithm}
\usepackage{multirow}
\usepackage{multicol}
\usepackage{algorithmic}
\usepackage{rotating}

\usepackage{graphics}
\usepackage{graphicx}

\usepackage{amssymb}
\usepackage{amsmath}
\usepackage{amsfonts}
\usepackage{calligra}
\usepackage{calrsfs}

\begin{document}




\title{Learning Weighted Association Rules in Human Phenotype Ontology.}

\author{Pietro Hiram Guzzi, Giuseppe Agapito, Marianna Milano, Mario Cannataro}
\maketitle 

\begin{abstract}
The Human Phenotype Ontology (HPO) is a structured repository of concepts (HPO Terms) that are associated to one or more diseases.  The process of association is referred to as annotation.  The relevance and the specificity of both HPO terms and annotations are evaluated by a measure defined as Information Content (IC). The analysis of annotated data is thus an important challenge for bioinformatics. There exist different approaches of analysis. From those, the use of Association Rules (AR) may provide useful knowledge, and it has been used in some applications, e.g. improving the quality of annotations. Nevertheless classical association rules algorithms do not take into account the source of annotation nor the importance yielding to the generation of candidate rules with low IC. This paper presents HPO-Miner (Human Phenotype Ontology-based Weighted Association Rules) a methodology for extracting Weighted Association Rules. HPO-Miner can extract relevant rules from a biological point of view. A case study on using of HPO-Miner on  publicly available HPO annotation datasets is used to demonstrate the effectiveness of our methodology.
\end{abstract}



\section{Introduction}
\label{sec:Intro}

In computer science, the term ontology defines a set of representational primitives with which to model a domain of knowledge or discourse \cite{gruber2009ontology}. In particular, ontologies are mainly used in bioinformatics and computational biology.

For instance, the Gene Ontology aims to provide a common language to describe genes product \cite{gene2004gene}. More recently, the annotation efforts have also focused on the description of relation among molecular biology and disease, leading to the introduction of novel ontologies such as Human Phenotype Ontology (HPO) \cite{} and Disease Ontology (DO) \cite{}.

HPO aims to provide a standardized vocabulary of phenotypic abnormalities encountered in human diseases. A generic HPO annotation contains a link between a disease and phenotypic abnormality. A disease is indexed by using a unified identifier known as Online Mendelian Inheritance in Man (OMIM). OMIM is a comprehensive, authoritative compendium of human genes and genetic phenotypes that are freely available and updated daily \cite{hamosh2005online}. The Disease Ontology (DO) has been developed as a standardized ontology for human disease with the purpose of providing strong and sustainable descriptions of human disease terms and phenotype characteristics \cite{schriml2012disease}.

The amount of annotations available is steadily growing, raising new challenges to face, related to ambiguous or incomplete annotations and ontology terms \cite{flouris2006inconsistencies}. The annotation task is becoming an even harder challenge in the genomic era, which is characterized by an unprecedented growth in the production of genes, gene products, and even other information. To speed-up the updating and maintenance processes of ontologies and annotations, it is required the development of computational approaches that guarantee a remarkable speed, on the current approaches of annotation carried out manually by the curators. 
The literature contains several computational methods developed to aid GO curators to improve GO annotations consistency \cite{yeh2003knowledge}, \cite{10.1371/journal.pone.0040519}, \cite{manda2012cross}. As opposed to GO, in literature, there are only a few automatic methodologies able to aid the HPO curators to improve annotation consistency and retrieve link between terms not explicitly related.  

As demonstrated in some recent works by Faria et al. \cite{faria2012}, by Manda et al. \cite{manda2013interestingness}, and by Agapito et al. \cite{agapito2014improving,agapito2015using},  association rules may be used to improve annotations consistency and highlight relationships among terms did not seem explicitly related. In this work, we present HPO-Miner an improvement of our previous works in which we introduced GO-WAR \cite{agapito2015using}. HPO-Miner is a tool for learning weighted association rules (WAR) to check annotation consistency and to identify hidden relationships between two phenotype abnormalities from HPO.
Traditional association rule approaches are not able to distinguish between items; they are unaware of the relevance of terms yielding to the generation of rules with low specificity. The specificity of each term may be measured by the information content (IC) of a term \cite{harispe2013framework}. The use of IC computed for each HPO term, is a measure of the specificity of a term, yielding to the IC-weighted annotation as conveyed in the following: \textit{OMIM100100: (HP:0000126, 11.18), (HP:0000144, 9.57)}.
HPO-Miner is able to extract weighted association rules starting from an annotated dataset of diseases. The proposed approach is based on the following steps: (i) initially we rearrange the information for each OMIM term to get transactional data; (ii) then, we extract weighted association rules using a modified \textit{FP-Tree} like algorithm able to deal with the dimension of classical biological datasets. We use publicly available HPO annotation data to demonstrate our method.

The rest of the paper is structured as follows:  Section \ref{sec:methods} discusses HPO-Miner methodology and implementation, Section \ref{sec:results} presents results of the application of HPO-Miner on a biological dataset. Finally Section \ref{sec:conclusion} concludes the paper.

\section{Materials and Methods}
\label{sec:methods}

\subsection{The Human Phenotype Ontology}

\begin{figure}[h]
\centering
 \includegraphics[width=4in]{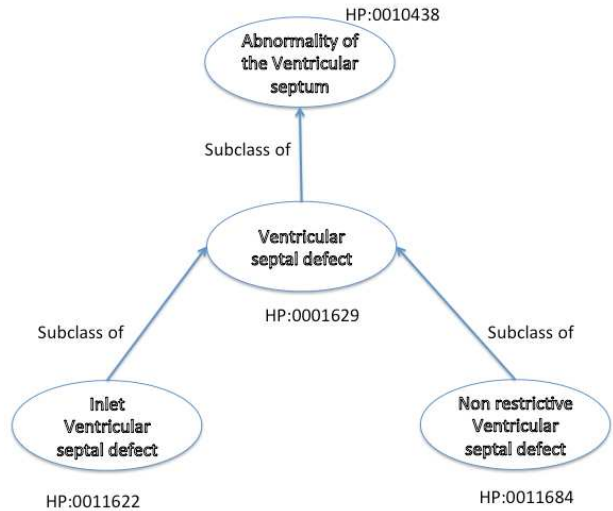}
   \caption{HPO graph Example}
\label{fig:path}
\end{figure}

 HPO is a structured and controlled vocabulary with more than 10,000 terms able to describe the phenotypic abnormalities in human diseases. HPO provides annotations of more than 7,000 human hereditary syndromes and other phenotypic abnormalities that characterize the diseases, are also available at the website \footnote{http://www.human-phenotype-ontology.org}. 
 HPO consists of three independent sub-ontologies: the \textit{mode of inheritance} i.e. the way in which a specific hereditary attribute is transmitted from a generation to another, \textit{onset and clinical course} i.e. in medicine refers to the first symptoms of a sickness and the medical treatments involved to cure them and finally, the \textit{phenotypic abnormalities} i.e.  the abnormal traits of a living organism that are possible to observe. As other ontologies, terms in HPO are organized in a directed acyclic graph (DAG). The relations among DAG's terms are modelled by means of \textit{is\_a} and \textit{part\_of} edges "relations", in order to distinguish between general or specific terms. Moreover, terms in HPO are arranged in a hierarchical way, where each path respects the \textit{true-path-rule}. To each HPO class is assigned a  stable and unique identifier (e.g. \textit{HP:0001629}), a label and a list of synonyms,  describing a well definite phenotypic abnormality i.e. "\textit{Ventricular Septal Defect}" see Figure \ref{fig:path}.

Diseases are annotated with terms of the HPO, meaning that HPO terms are used to describe all the signs, symptoms, and other phenotypic manifestations that characterize the disease in question. 
 
The annotations of OMIM entries are a mixture of manual annotations performed by the HPO curators team and automated matching of the OMIM Clinical Synopsis to HPO term labels. In particular HPO is an ontology designed to provide qualitative information and not to capture quantitative information such as body weight or height. Each diseases may be annotated to multiple HPO terms. 
Consequently the need of the introduction of methodologies and tools to support HPO curators to improve annotation consistency and the structure of the ontology arises. 

\subsection{Association Rules}

Association Rule (AR) extraction is very popular in data mining, it is used for discovering associations in market basket analysis and unknown relations among features in databases. Historically, was proposed by Agrawal \cite{citeulike:1005421} to discovery associations to support marketing decision. 
 
Formally, the association rules extraction problem may be stated as follows: let $ I=\{i_1,i_2,\ldots,i_{n}\}$ be a set of items 
and $D = \{t_1, \ldots, t_m \}$ a transactional database that contains a set of transactions, where 
 a transaction $t_j$ is a subset of items belonging to $I$. An association rule is an implication of the form $A \rightarrow B$, where $A$ and $B$ are two disjoint sets. 
 AR are based on two fundamental properties to define the relevance of the mined rules, \textit{Support} and \textit{Confidence}.
The formal Support definition is: 
\begin{definition}
\centering	
	$S(A \rightarrow B) = \frac{  \sigma (A \cup B)}{N} $
\end{definition}
Where $N$ is the total number of transactions contained in $D$ and $\sigma$ is called \textit{support count}, namely, the number of transaction that contain a particular item.
\newline The Confidence is defined as: 
\begin{definition}
\centering	
	$C(A \rightarrow B) = \frac{  \sigma (A \cup B)}{\sigma (A)} $.
\end{definition}

Where $\sigma(A)$ is the number of transactions in \textit{D} containing A and $\sigma(A\cup B)$ is the number of transactions in \textit{D} that contains both items A and B.\newline

A drawback with the use of classical AR approach is that it precludes the derivation of certain rules in which the items have a very different levels of support. In several areas do not make sense to assign equal importance to all items involved in the dataset. For example in the supermarket context, some items like computer, smartphone have much  value than trivial items like ice-cream or butter. Rules involving smartphone or computer have less support than those involving butter or ice-cream but are much more significant in term of profit by the store. In the ontology context, the term HP:0000924 (\textit{An abnormality of the skeletal system}) has a relevance value (IC value) lower than HP:0011803 (\textit{Bifid nose}) although it is much more frequent. Rules involving the term HP:0000924 are less interesting (as it is a more general term) then rules involving the term HP:0011803 (as it is a more specific term) in terms of actionable knowledge. 

This limitation of classical AR approach can be overcome by introducing the weighted association rules (WAR). WAR models the \textit{significance} of a term by means of a \textit{weight} ($\omega$). A weight ($\omega$) is a positive real number that reflect the relevance of a HPO terms,  for which high values represent very significant items as reported in \cite{Wang:2000:EMW:347090.347149, 694360}. In our case, the relevance can be represented by using the information content (\textit{IC}). 

\begin{figure}[h]
\centering
 \includegraphics[width=2.0in]{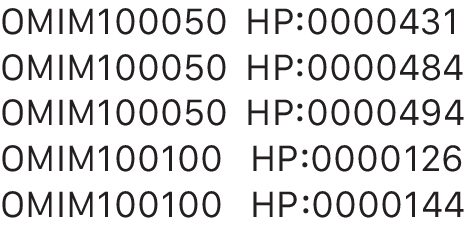}
   \caption{An example of HPO dataset.}
\label{fig:HPO_daataset}
\end{figure}

A generic HPO dataset is a list of \textit{OMIM} identifiers annotated with multiple HPO terms, as conveyed in Figure \ref{fig:HPO_daataset}.

\begin{figure}[h]
\centering
 \includegraphics[width=4.0in]{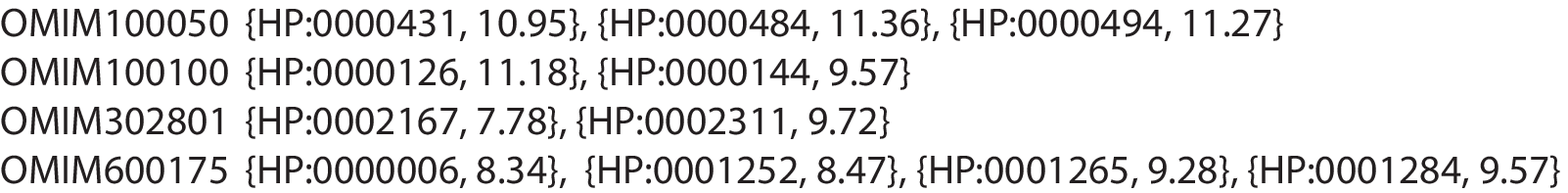}
   \caption{An example of weighted transaction HPO dataset.}
\label{fig:WD}
\end{figure}

In order to extract rules from the HPO dataset, it is necessary to convert it in a format more suitable to represent transaction data. The conversion consists in put together the same OMIM identifiers that became the transaction \textit{identifier} while the HPO terms associated with the current OMIM identifier are the items of the transaction, as depicted in Figure \ref{fig:WD}. 



\subsection{Weighting HPO term with Information Content}
Each HP term is associated to IC value. There exist different IC formulations  that fall into two classes, intrinsic and extrinsic methods. Intrinsic method rely on the topology of the GO graph analyzing the positions of terms in a taxonomy. In this way the approaches define information content for each term. Different topological characteristics as ancestors, number of children, depth (see\cite{harispe2013framework} for a complete review) can used in order to estimate the Intrinsic IC calculus. Instead the extrinsic approaches involve annotation data for an considered corpus.
In this work we used the intrinsic method proposed by Sanchez et al. \cite{sanchez2011ontology}, Harispe et al.\cite{harispe2013framework}, Resnick et al. \cite{resnink:simmeasure:879855}, Seco et al.  \cite{14755292}, Zhou et al. \cite{zhou2008new}. 

The measure of Sanchez exploits only the number of leaves and the set of  ancestors of $a$ including itself, \emph{subsumers(a)} and introduce the root node as the number of leaves \emph{max\_leaves} in IC assessment. Leaves are more informative than concepts with many leaves, roots, so the leaves are suited  to describe and to distinguish any concept.
\begin{equation}
IC_{Sanchez\,et\,al.}(a) =-log\left(\frac{\frac{|leaves(a)|}{|subsumers(a)|}+1}{max\_leaves+1} \right)
\end{equation}

Harispe et al., in oder to highlights the specificity of leaves according to their number of ancestors,   consider \emph{leaves(a)} = \emph{a} concept when \emph{a} is a root and evaluating \emph{max\_leaves} as the number of inclusive ancestors of a node revising  the IC assessment  suggested by Sanchez et al.
\begin{equation}
IC_{Harispe\,et\,al.}(a) =-log\left(\frac{\frac{|leaves(a)|}{|subsumers(a)|}}{max\_leaves} \right)
\end{equation}
 
  The  formulation provided from Resnick et al. computes the IC of a concept  evaluating all the top-downs path from a concept $a$ to the reachable leaves, $p(a)$, and then calculates the log  yielding to the formula: \begin{equation}
IC_{Resnik}(a) = -log(p(a)).
 \end{equation}

Seco et al. calculate the IC of a concept by considering the ratio between the number of hyponyms in ontology, for example, the number of descendant with respect to the whole  number of ontological concepts.
\begin{equation}
{IC_{Seco\,et\,al}(a) =\frac{log\left( \frac{hypo(a)+1}{max\_nodes} \right)}{log\left( \frac{1}{max\_nodes} \right)} }
\end{equation}

 Thus Zhou et al. considers the depth of a concept in a taxonomy, \emph{depth(a)}, and the maximum depth of the taxonomy  \emph{max\_depth}.
\begin{equation}
\small{IC_{Zhou\,et\,al.}(a) =k-\left(1- \frac{log(hypo(a)+1)}{log(max\_nodes)} \right)+(1-k)\left(\frac{log(depth(a))}{log(depth\_nodes)} \right)}
\end{equation}
In this formulation K is a factor which enables to weight the contribution of the two evaluated features.

\section{The HPO-Miner Algorithm}
\label{sec:HPO-Miner}

In this section we briefly describe the \textit{HPO-Miner} algorithm, developed to extract weighted association rules form HPO dataset. 

First of all we define the \textit{Weighted Item $x$}, i.e. a weighted HPO item is obtained by multiplying the number of occurrences of item $x$ by the value of its related value of IC (the weight $\omega$). We define as $Weighted Support$, ($\omega S$), obtained by integrating  the classical formulation of the support of an item by its weight. The weighted Support \textit{$\omega S$} of a generic item $x_i$ is defined as: $\omega S(x_i)=w_i*\sigma(x_i)$ where $\omega _i$ is the information content of the \textit{i-th} term and $\sigma(x_i)$ is the number of transaction containing $x_i$. Let $I=\{i_1 \ldots i_m\}$ be a set of weighted items (HPO terms) and let $WD$ be a set of weighted transactions database, where each transaction $t_j$ is a sub-set of weighted items such that $t_j$ belongs to $I$. 
 We defined the \textit{weighted minimum support ($m\omega S$)} as:\\
\begin{definition}
\centering	
$m\omega S =  \left(\frac{\sum_{i=1}^{|WD|} \sigma(x_i)*\omega_i} {|WD|}\right)*p$.
 \end{definition}
 Where, $|WD|$ is the cardinality of the weighted database nominally, the number of transactions into the dataset, $p$ is a threshold value given in input by the user in order to define which items are significant in percentage.  Thus only the items for which the following constraint $\omega S(I) \geq m\omega S$ is verified, are significant and can be used as candidates to generate frequent item-sets and rules.

Algorithm \ref{alg:mapping} is a summary of the main phases of the\textit{HPO-Miner} algorithm.
The first step of HPO-Miner algorithm is the loading of the input HPO dataset ($D$) and its transformation in \textit{weightedTable} $WT$ a data structure suitable to represent weighted transaction data (as reported in Algorithm \ref{alg:mapping} row 2). Concurrently to the loading and conversion phase, are evaluated the occurrences of each HPO term in $D$. Subsequently is possible to obtain a list of frequent weighted items (as stated in Algorithm \ref{alg:mapping} at row 3). We remove from the $\mathcal{FW}ItemsList$ the weighted items for which is not verified the following condition: $\omega S(I) \geq m\omega S$. Frequent weighted items are hence used to build a data structure based on  $FP-Tree$.  Finally, \textit{HPO-Miner} iteratively analyzes the $FP-Tree$ in order to mine and save significant rules. 
\begin{algorithm}
\caption{HPO Weighted Association Rules Miner (HPO-Miner)}\label{alg:mapping}
\label{alg:mapping}
\begin{algorithmic}[1]
   \small \REQUIRE A table of HPO annotation as input dataset $D$
    \STATE \textbf{\textit{Data Structure initialization: $WT$, $\mathcal{FW}ItemsList$, \textit{FPTree}}}
    \medskip
    \STATE $WT \leftarrow $\textit{getTransactionalData($D$)}
    \medskip
	\STATE $\mathcal{FW}ItemsList$ $\leftarrow$ retrieve$\mathcal{FW}ItemsList$($WT$)
   \medskip
       \STATE{$FPTree.$create($\mathcal{FW}ItemsList$) }
        \medskip
       \medskip
       \STATE {$mineWeightedRules()$}
       \STATE \textbf{end.}

\end{algorithmic}
\end{algorithm}



%

\section{Results}
\label{sec:results}

\texttt{HPO} database is freely available online \footnote{\url{http://www.human-phenotype-ontology.org/downloads.html}}, the size of the dataset is about $4.4$ \texttt{MB} on disk. After collecting data, by using all the methods introduced in Section \ref{sec:methods}, we produced $5$ different datasets. 
We tested HPO-Miner using several combinations of values for weightedSupport and confidence. Then we selected the values for the parameters able to ensure the best results in terms of reduced number of mined rules and in the same time with relevant values of weightedSupport and confidence. The best combination of values was weightedSupport equal to $50\%$ and confidence greater than $80\%$. We chose the first top $10$ rules from each dataset, and we manually analyzed the literature to find claims that can prove the validity of the mined rules. 

\subsection{HPO-Miner rules extraction comparison}
The effectiveness of \textit{HPO-Miner} is proved comparing our tool with respect to other well known tools such as: Knime \cite{citeulike:5770121} and Weka \cite{weka}. We chose these tools because both provides an implementation of the FP-Growth algorithm a necessary condition in order to fairly compare HPO-Miner with both tools.  The FP-Growth algorithm implementation in Weka and Knime, is able to handle only binary attributes, making both tools unable to analyze weighted HPO datasets enriched with IC values. A possible way to make weighted HPO enriched dataset compatible with Weka and Knime is to leave for each OMIM entry only two HPO terms, making this solution infeasible because leads to lose a lot of useful information. 
Differently, HPO-Miner is the only tool that comes with a version of FP-Growth able to handle a generic number of attribute for each OMIM entry, making it suitable to analyze HPO dataset enriched with IC values.

\subsection{Analysis of Mined Rules}


\begin{table}[ht]

\caption{The ten first rules found by HPO-Miner using the Dataset obtained by applying the Resnik measure and ranked by weightedSupport. (IDs are inserted for a better discussion in the following.) }
\label{tab:Resnik}
\begin{tabular}{|p{0.5cm}|l|l|p{0.5cm}|p{0.5cm}|p{2.0cm}|p{2.0cm}|}
\hline 
\textbf{}&\textbf{Term 1}& \textbf{Term 2} & \textbf{WS} & \textbf{C} &\textbf{Function} & \textbf{Function }\\ 
\hline
$1R$ & HP:0200084 & HP:0000007 &$1.00$&$1.00$&Giant cell hepatitis&Autosomal recessive inheritance\\
\hline
$2R$&HP:0200084&HP:0002910&$1.00$&$1.00$&Giant cell hepatitis&Elevated hepatic transaminases\\
\hline
$3R$&HP:0200067&HP:0000006&$1.00$&$1.00$&Recurrent spontaneous abortion&Autosomal dominant inheritance\\
\hline
$4R$&HP:0100818&HP:0000774&$1.00$&$1.00$&Long thorax&Narrow chest\\
\hline
$5R$&HP:0100775&HP:0001537&$1.00$&$1.00$&Dural ectasia&Umbilical hernia\\
\hline
$6R$&HP:0100775&HP:0000006&$1.00$&$1.00$&Dural ectasia&Autosomal dominant inheritance\\
\hline
$7R$&HP:0100775&HP:0000494&$1.00$&$1.00$&Dural ectasia&Downslanted palpebral fissures\\
\hline
$8R$&HP:0100775&HP:0000316&$1.00$&$1.00$&Dural ectasia&Hypertelorism\\
\hline
$9R$&HP:0100626&HP:0001394&$1.00$&$1.00$&Chronic hepatic failur&Cirrhosis\\
\hline
$10R$&HP:0100626&HP:0000007&$1.00$&$1.00$&Chronic hepatic failure&Autosomal recessive inheritance\\
\hline
\end{tabular}
\end{table}

\begin{table}[ht]

\caption{The ten first rules found by HPO-Miner using the Dataset obtained by applying the Sanchez measure and ranked by weightedSupport. (IDs are inserted for a better discussion in the following.) }
\label{tab:Sanchez}
\begin{tabular}{|p{0.5cm}|l|l|p{0.5cm}|p{0.5cm}|p{2.2cm}|p{2.2cm}|}
\hline 
\textbf{}&\textbf{Term 1}& \textbf{Term 2} & \textbf{WS} & \textbf{C} &\textbf{Function} & \textbf{Function }\\ 
\hline
$1S$&HP:0100818&HP:0000774&$0.88$&$1.00$& Long thorax&Narrow chest\\
\hline
$2S$&HP:0030034&HP:0003774&$0.88$&$1.00$&Diffuse glomerular basement membrane lamellation&Stage 5 chronic kidney disease\\
 \hline
$3S$&HP:0012743	&HP:0001773&$0.88$&$1.00$&Abdominal obesity&Short foot\\ 
 \hline
$4S$&HP:0012263&HP:0000007&$0.88$&$1.00$&Immotile cilia&Autosomal recessive inheritance\\
 \hline
$5S$&HP:0012023&HP:0000007&$0.88$&$1.00$&Galactosuria&Autosomal recessive inheritance\\ 
 \hline
$6S$&HP:0011727&HP:0009049&$0.88$&$1.00$&Peroneal muscle weakness&Peroneal muscle atrophy\\
 \hline
$7S$&HP:0010636&HP:0000316&$0.88$&$1.00$&Schizencephaly&Hypertelorism\\
 \hline
$8S$&HP:0009793&HP:0000316&$0.88$&$1.00$& Presacral teratoma&Hypertelorism\\
 \hline
$9S$&HP:0009760&HP:0006443&$0.88$&$1.00$&Antecubital pterygium&Patellar aplasia\\
 \hline
$10S$&HP:0008845&HP:0003067&$0.88$&$1.00$&Mesomelic short stature&Madelung deformity\\
\hline
\end{tabular}
\end{table}

\begin{table}[ht]

\caption{The ten first rules found by HPO-Miner using the Dataset obtained by applying the Harispe measure and ranked by weightedSupport. (IDs are inserted for a better discussion in the following.) }
\label{tab:Harispe}
\begin{tabular}{|p{0.5cm}|l|l|p{0.5cm}|p{0.5cm}|p{2.0cm}|p{2.0cm}|}
\hline 
\textbf{}&\textbf{Term 1}& \textbf{Term 2} & \textbf{WS} & \textbf{C} &\textbf{Function} & \textbf{Function }\\ 
\hline
$	1H	$&	HP:0009577	&	HP:0004220	&$	1.00	$&$	1.00$&	Short middle phalanx of the 2nd finger   &      Short middle phalanx of the 5th finger \\ 
\hline 
$	2H	$&	HP:0010105	&	HP:0010034	&$	1.00	$&$	1.00	$&	Short first metatarsal      &     Short 1st metacarpal\\ 
\hline
$	3H	$&	HP:0000933	&	HP:0001305	&$	1.00	$&$	1.00	$&	Posterior fossa cyst at the fourth ventricle	 &   Dandy-Walker malformation\\ 
\hline
$	4H	$&	HP:0004704	&	HP:0004689	&$	1.00	$&$	1.00	$&	Short fifth metatarsal	&      Short fourth metatarsal\\ 
\hline
$	5H	$&	HP:0001885	&	HP:0004209	&$	1.00	$&$	0.99	$&	Short 2nd toe   &   Clinodactyly of the 5th finger\\ 
\hline
$	6H	$&	HP:0003065	&	HP:0006443	&$	1.00	$&$	1.00	$&	Patellar hypoplasia  &     Patellar aplasia\\ 
\hline
$	7H	$&	HP:0009464	&	HP:0004209	&$	1.00	$&$	1.00	$&	 Ulnar deviation of the 2nd finger   &    Clinodactyly of the 5th finger\\ 
\hline
$	8H	$&	HP:0002834	&	HP:0002857	&$	1.00	$&$	1.00	$&	Flared femoral metaphysis     &    Genu valgum\\ 
\hline
$	9H	$&	HP:0004209	&	HP:0000272	&$	1.00	$&$	0.99	$&	Clinodactyly of the 5th finger &    Malar flattening\\ 
\hline
$	10H	$&	HP:0001773	&	HP:0004279	&$	1.00	$&$	1.00	$&	  Short foot   &  Short palm\\ 
\hline
\end{tabular}
\end{table}

\begin{table}[ht]

\caption{The ten first rules found by HPO-Miner using the Dataset obtained by applying the Seco measure and ranked by weightedSupport. (IDs are inserted for a better discussion in the following.) }
\label{tab:Seco}
\begin{tabular}{|p{0.55cm}|l|l|p{0.5cm}|p{0.5cm}|p{2.0cm}|p{2.0cm}|}
\hline 
\textbf{}&\textbf{Term 1}& \textbf{Term 2} & \textbf{WS} & \textbf{C} &\textbf{Function} & \textbf{Function }\\ 
\hline
$	1Se	$&	HP:0200084	&	HP:0000007	&$	1	.00$&$	1	.00$&	Giant cell hepatitis	&	Autosomal recessevie inheritance\\ 
\hline
$	2Se	$&	HP:0100818	&	HP:0000774	&$	1	.00$&$	1	.00$&	Long thorax	&	Narrow chest\\ 
\hline
$	3Se	$&	HP:0100775	&	HP:0001537	&$	1	.00$&$	1	.00$&	Dural ectasia	&	humbilical hernia\\ 
\hline
$	4Se	$&	HP:0100775	&	HP:0000494	&$	1	.00$&$	1	.00$&	Dural ectasia	&	Downslanted palpebral fissures\\ 
\hline
$	5Se	$&	HP:0100775	&	HP:0000316	&$	1	.00$&$	1	.00$&	Dural ectasia	&	Hypertelorism\\ 
\hline
$	6Se	$&	HP:0100626	&	HP:0000007	&$	1	.00$&$	1	.00$&	Chronic hepatic failure	&	Autosomal recessevie inheritance\\ 
\hline
$	7Se	$&	HP:0030050	&	HP:0002524	&$	1	.00$&$	1	.00$&	Narcolepsy	&	Cataplexy\\ 
\hline
$	8Se	$&	HP:0012240	&	HP:0000007	&$	1	.00$&$	1	.00$&	Increased intramyocellular lipid droplets	&	Autosomal recessevie inheritance\\ 
\hline
$	9Se	$&	HP:0010780	&	HP:0007018	&$	1	.00$&$	1	.00$&	Hyperacusis	&	Attention deficit hyperactivity disorder (ADHD)\\ 
\hline
$	10Se	$&	HP:0010780	&	HP:0000179	&$	1	.00$&$	1	.00$&	Short 3rd metacarpal	&	Hypertelorism\\ 
\hline
\end{tabular}
\end{table}

\begin{table}[!h]

\caption{The ten first rules found by HPO-Miner using the Dataset obtained by applying the Zhou measure and ranked by weightedSupport. (IDs are inserted for a better discussion in the following.) }
\label{tab:Zhou}
\begin{tabular}{|p{0.55cm}|l|l|p{0.5cm}|p{0.5cm}|p{2.0cm}|p{2.0cm}|}
\hline 
\textbf{}&\textbf{Term 1}& \textbf{Term 2} & \textbf{WS} & \textbf{C} &\textbf{Function} & \textbf{Function }\\ 
\hline
$1Z$&	HP:0002335	&	HP:0001305	&$	0.97	$&$	1	$&	Congenital absence of the vermis of cerebellum	&	Dandy Walker malformation\\ 
\hline
$2Z$&	HP:0003031	&	HP:0002986	&$	0.95	$&$	1	$&	Bending of the diaphysis (shaft) of the ulna (Uknar bowing)	&	A bending or abnormal curvature of the radius (Radial bowing)\\ 
\hline
$3Z$&	HP:0000176	&	HP:0000193	&$	0.95	$&$	0.97	$&	submucous clefts Hard-palate	&	Bifid uvula\\ 
\hline
$4Z$&	HP:0001338	&	HP:0002007	&$	0.95	$&$	0.94	$&	Partial agenesis of the corpus callosum	&	Frontal Bossing\\ 
\hline
$5Z$&	HP:0001338	&	HP:0000494	&$	0.95	$&$	0.94	$&	Partial agenesis of the corpus callosum	&	Downslated palpebral fissures\\ 
\hline
$6Z$&	HP:0000308	&	HP:0001305	&$	0.95	$&$	1	$&	Microre~trognathia	&	Dandy Walker malformation\\ 
\hline
$7Z$&	HP:0000269	&	HP:0001305	&$	0.95	$&$	1	$&	Promiment occiput	&	Dandy Walker malformation\\ 
\hline
$8Z$&	HP:0010804	&	HP:0001305	&$	0.95	$&$	1	$&	Tented upper lip vermilion	&	Dandy Walker malformation\\ 
\hline
$9Z$&	HP:0009623	&	HP:0001305	&$	0.95	$&$	1	$&	Proximal placement of the thumb	&	Dandy Walker malformation\\ 
\hline
$10Z$&	HP:0000567	&	HP:0001305	&$	0.95	$&$	1	$&	Chorioretinal coloboma	&	Dandy Walker malformation\\ 
\hline
\end{tabular}
\end{table}


Let us consider rule (1R): (HP:0200084, HP:0000007) - \textit{Giant cell hepatitis, Autosomal recessive inheritance}. Searching the literature we found some evidences that describe the relationship between this two terms. As stated in \cite{Danks01051977} both terms could be related with defects in the biological mechanisms of the liver. In particular,
\textit{Autosomal recessive inheritance} suggests a biochemical defect that might cause a metabolic disorder in the liver while, \textit{Giant cell hepatitis} is responsible of "thick bile syndrome" in neonatal. Consequently, HPO-Miner was able to found a relation between two apparently unrelated terms into the graph of HPO classes.

Rule (2R) (HP:0200084, HP:0002910) i.e.,  (\textit{Giant cell hepatitis}, \textit{Elevated hepatic transaminases}) consists of two terms involved in the hepatitis process. Analyzing in depth the literature it revealed   the following links between the two terms. In \cite{APA:APA285} is presented a study on three siblings with neonatal jaundice who died before the age of three months. They were shown on autopsy to be suffering from Niemann-Pick disease together with a giant cell transformation of the liver. Clayton et. al. in \cite{ClaytonPaper} including the infant studied in \cite{Kase1985} were able to inferrer, that due to the elevated transaminases most patients develop hepatic fibrosis or cirrhosis due to the presence of \textit{Giant cell hepatitis}. Thus, manually analyzing this rule has been possible to infer that both terms are responsible of the liver disorder in infants and adults.

Rule (3R) involves the following two HPO terms (HP:0200067, HP:0000006) i.e.,\textit{Recurrent spontaneous abortion} and \textit{Autosomal dominant inheritance}. There is a growing literature on the importance of  Autosomal dominant inheritance in pregnancy complications as reported in \cite{byrne1994genetic}. As stated in \cite{Kutteh2006} Thrombophilia is a cause of maternal mortality due to certain inherited thrombophilic factors that activated protein C resistance. In \cite{coumans1999haemostatic} the authors point out the rare familial disorders that are usually inherited as \textit{Autosomal dominant inheritance}. 

Rule (4R) (HP:0100818, HP:0000774) composed by the following phenotypic abnormalities \textit{Long thorax}, \textit{Narrow chest} involved in the syndrome of Jeune and Ellis-Van Creveld syndrome as reported in literature in \cite{AJMG:AJMG1320210304, baujat2007ellis}. Browsing HPO Ontology with its on line browser did not reveal any information that allows the user to associate both abnormalities with the syndrome of Juene and Ellis-Van Creveld. This may suggest to the curator to restructure ontology in order to make easily available this knowledge in order to clarify these associations.

Rule (5R) (HP:0100775, HP:0001537) whose translation is \textit{Dural ectasia}, \textit{Umbilical hernia} at first glance seems that there not exists a connection among the two terms. Analyzing the literature we found the work of Mizuguchi et.al. \cite{mizuguchi2004heterozygous} and Chen et. al., \cite{chen2005lateral}. In Mizuguchi et.al. have been found both abnormalities in a patient affected by the Marfan syndrome in infancy, instead Chen et. al. have found  these abnormalities in patients affected by Lateral meningocele syndrome. These knowledge it is not readily available for the users by using HPO, consequently this may suggest to the curator to add this further knowledge into the HPO. 

Rule (6R) (HP:0012023, HP:0000007) define an association between the   \textit{Galactosuria} and \textit{Autosomal recessive inheritance}. Analyzing the literature looking for evidence on the validity of the association we found the works of Pickering et. al., \cite{pickering1972galactokinase} and Monteleone et. al. \cite{monteleone1971cataracts}, in which in both works, the authors stated that hereditary galactokinase deficiency is characterized by galactosuria. In particular, in this study support the autosomal recessive inheritance of this disorder. This evidence support the validity of the current association found it by using HPO-Miner.

To verify the reliability of Rule (7R) (HP:0100775,HP:0000494) i.e. (\textit{Dural ectasia}, \textit{Downslanted palpebral fissures}) and Rule (8R) (HP:0100775, HP:0000316) i.e., \textit{Dural ectasia}, \textit{Hypertelorism}, we analyzed the literature founding that the terms of both rules are symptoms involved in the Marfan syndrome as stated in \cite{lemaire2007severe, loeys2010revised}. Consequently these association rules may suggest to the curator to add new informative links among HPO terms, making easier for the users to obtain further knowledge.

Rule (9R) (HP:0100626, HP:0001394) refers to \textit{Chronic hepatic failure} and \textit{Cirrhosis}. Analyzing the literature showed that both terms are involved in fat elimination as stated in the work of Druml et. al. \cite{druml1995fat}. This evidence may be suggest to the curator to make this explicit knowledge in implicit, by adding new links among the HPO terms.

 (10R) (HP:0100626, HP:0000007) \textit{Chronic hepatic failure}, \textit{Autosomal recessive inheritance}

Here we discuss the rules contained in Table \ref{tab:Sanchez} that refer to the rules mined by HPO-Miner from the Sanchez dataset.

Rule (1S) (HP:010081, HP:0000774) i.e.,  (\textit{Long thorax}, \textit{Narrow chest}) consists in two terms involved in the Asphyxiating Thoracic Dysplasia (Jeune Syndrome).Jeune syndrome is a congenital disorder with abnormalities of which thoracic hypoplasia is the most prominent. The literature confirms that both phenotype, long thorax and narrow chest are manifestations of Jeune syndrome. In \cite{elejalde1985prenatal} is reported this evidence.

Rule (2S) (HP:0030034, HP:0003774) associates  with\textit{Diffuse glomerular basement membrane lamellation}, \textit{Stage 5 chronic kidney disease}. Searching in the current literature the glomerular basement membrane lamellation is a manifestation in patients after transplantation of kidneys from pediatric cadaveric donors, as \cite{nadasdy1999diffuse} reported. There is not evidence that this phenotype is related to the Stage 5 chronic kidney disease.

About the Rule (3S) (HP:0012743, HP:0001773), \textit{Abdominal obesity}, \textit{Short foot} we didn't find a correlation among \textit{Abdominal obesity}(term 1) and \textit{Short foot} (term 2) despite a depth research in literature was conducted .

Rule (4S) (HP:0012263, HP:0000007) and Rule (5S) (HP:0012023, HP:0000007), associate two pathologic phenotypes, \textit{Immotile cilia} and (\textit{Galactosuria} to \textit{Autosomal recessive inheritance}). In fact, in \cite{afzelius1985immotile} is reported that the immotile cilia syndrome seems to be that of an autosomal recessive disease; as well as   galactosuria due to galactokinase deficiency in a newborn is inherited in an autosomal recessive manner \cite{pickering1972galactokinase}.

HPO-MINER finds the Rule (6S) (HP:0011727, HP:0009049) that associates(\textit{Peroneal muscle weakness} with \textit{Peroneal muscle atrophy}).In fact the peroneal muscle atrophy is characterized
by wasting and flaccid weakness of the intrinsic muscles of the feet and of the muscles innervated by the peroneal nerve \cite{buchthal1977peroneal}.

Rule (7S) (HP:0010636, HP:0000316) relates (\textit{Schizencephaly}, \textit{Hypertelorism}) involved in the same disease, the LEOPARD syndrome. A case study \cite{liang2009schizencephaly} reported patient affect by this disease with open-lip schizencephaly and Ocular hypertelorism pathologic phenotype.

Instead Rule (8S) (HP:0009793, HP:0000316), highlights a link among \textit{Hypertelorism}) with (\textit{Presacral teratoma} in the Schinzel‐Giedion syndrome as reported in \cite{robin1993new}.

In \cite{reichenbach1995hereditary} is discussed a Hereditary Congenital Posterior Dislocation
of Radial Heads which disorder is characterized by The association of nailpatella syndrome
with typical antecubital pterygium as HPO-MINER found in Rule (9S) (HP:0009760, HP:0006443), 

Rule (10S) (HP:0008845, HP:0003067), composed by (\textit{Mesomelic short stature}, \textit{Madelung deformity}). Both phenotype are involved in Madelung deformity of childhood \cite{flanagan2002prevalence}

Here we analyze the rules contained in Table \ref{tab:Seco}.

Rule (1Se) (HP:0200084, HP:0000007) associates \textit{Giant cell hepatitis}and  \textit{Autosomal recessive inheritance}. This evidence is highlighted in a case study reported a patient suffered from a unique form of giant cell hepatitis which condition appears to be an autosomal recessive one\cite{clayton1987familial}

Rule (2Se) (HP:0100818, HP:0000774) \textit{Long thorax}, \textit{Narrow chest} is discussed above.

About the Rule (3Se) (HP:0100775, HP:0001537) i.e. \textit{Dural ectasia}, \textit{humbilical hernia} HPO-MINER find a association that is not confirmed in literature. 

The Rule (4Se) (HP:0100775, HP:0000494) and the Rule (5Se) (HP:0100775, HP:0000316) associate the phenotype\textit{Dural ectasia} with\textit{Downslanted palpebral fissures} and \textit{Hypertelorism}. Carrying out a analysis in the state of art, we found a clinical case which report a patient with lateral meningocele syndrome (LMS) affected by both down slanting palpebral fissures and hyperteloris \cite{chen2005lateral}.

About the Rule (6Se) (HP:0100626, HP:0000007) associates \textit{Chronic hepatic failure} to characteristic \textit{Autosomal recessive inheritance}
\cite{blumberg1969hepatitis}

In the Rule (7Se) (HP:0030050, HP:0002524) are connected two pathologic phenotype \textit{Narcolepsy} and \textit{Cataplexy} that are known as a sleep disorder associated with a centrally mediated hypocretin deficiency\cite{mignot2001complex}.

About Rule (8Se) (HP:0012240, HP:0000007) the evidence that the \textit{Increased intramyocellular lipid droplets} is \textit{Autosomal recessive inheritance}.

The Rule (9Se) (HP:0010780, HP:0007018) associates the symptom  \textit{Hyperacusis} to \textit{Attention deficit hyperactivity disorder (ADHD)} as reported in\cite{einfeld1995issues}

Instead the Rule (10Se) (HP:0000179, HP:0010780) i.e\textit{Short 3rd metacarpal}, \textit{Hyperacusis}  has not evidence in literature. 

Here we interpret the rule mined by HPO-Miner from the dataset Harispe and contained in Table \ref{tab:Harispe}.

Rule (1H) is composed of terms (HP:0009577, HP:0004220)	i.e., (Short middle phalanx of the 2nd finger, Short middle phalanx of the 5th finger). Analyzing the literature, we found that this abnormalities have been observed in the Adams-Oliver Syndrome as reported in the work of Kuster et.al. \cite{kuster1988congenital}.

Rule (2H) contains the terms	(HP:0010105, HP:0010034) i.e., \textit{Short first metatarsal, Short 1st metacarpal}

Rule (3H) (HP:0000933,HP:0001305) i.e., \textit{Posterior fossa cyst at the fourth ventricle} \textit{Dandy-Walker malformation} involved in abnormality that affects brain development.

Analyzing the literature has not been possible found any evidence on the involvement of the (HP:0004704, HP:0004689)	i.e., \textit{Short fifth metatarsal, Short fourth metatarsalRule}, contained in the rule (4H) found by HPO-Miner.

Rule (5H) is formed by the two terms (HP:0001885, HP:0004209)	i.e., \textit{Short 2nd toe, Clinodactyly of the 5th finger}. Searching into the literature we found that both symptoms occurred in  Carpenter Syndromeas states in the work of Gershoni et.al., \cite{gershoni1990carpenter}.

Rule (6H) involves the following two HPO terms (HP:0003065, HP:0006443) 	i.e., \textit{Patellar hypoplasia,  Patellar aplasia}. The work of Kaariainen et. al. \cite{kaariainen1989rapadilino} that RAPADILINO syndrome involve both symptoms.

Rule (7H) (HP:0009464, HP:0004209) i.e., \textit{Ulnar deviation of the 2nd finger, Clinodactyly of the 5th finger} consists of two terms involved in the KBG syndrome as reported in the work of Sirmaci et. al. \cite{sirmaci2011mutations}.

Rule (8H) is composed of (HP:0002834, HP:0002857)	i.e., \textit{Flared femoral metaphysis, Genu valgum}. Both symptom are observed in the metatropic dwarfism as described into the work of LaRose et. al. \cite{larose1969metatropic}.

The terms contained into the rule (9H) HP:0004209, HP:0000272 i.e.,	\textit{Clinodactyly of the 5th finger, Malar flattening} are involved in \textit{49,XXXXY} syndrome as stated in the work of Peet et. al. \cite{peet199849}.

About Rules (10H) (HP:0001773, HP:0004279) i.e. \textit{Short foot, Short palm} we didn't find any correlation between the terms, despite a depth research in literature it was conducted.

\pagebreak
Here we analyze the rules contained in Table \ref{tab:Zhou}.

The first rule Rule (1Z) (HP:0002335, HP:0001305) associates the \textit{Congenital absence of the vermis of cerebellum} with \textit{Dandy Walker malformation}. This evidence is confirmed in \cite{bordarier1990dandy}, that reported a cases of Dandy-Walker malformation including agenesis cerebellar vermis.

HPO-MINER extracts the Rule (2Z) (HP:0003031, HP:0002986) i.e\textit{Bending of the diaphysis (shaft) of the ulna (Ulnar bowing)}  \textit{A bending or abnormal curvature of the radius (Radial bowing)} and the Rule (3Z) (HP:0000176, HP:0000193), i.e.\textit{submucous clefts Hard-palate	} \textit{Bifid uvula}. Although we conducted a deep analysis of stare of art, these rules are not confirmed in literature.

The Rule (4Z) (HP0001338 HP0002007) and the Rule (5Z) (HP0001338 HP0000494) associate the \textit{Partial agenesis of the corpus callosum} to two abnormal phenotype: \textit{Frontal Bossing} and \textit{Downslated palpebral fissures} as confirmed in \cite{taylor1968nevoid} and\cite{gelman1991further}.

HPO-MINER finds the Rule (6Z) (HP:0000308, HP:0001305), Rule (8Z) (HP:0010804, HP:0001305), Rule (9Z) (HP:0009623, HP:0001305) that associate the phenotypes\textit{Microretrognathia} \textit{Tented upper lip vermilion} in, \textit{Proximal placement of the thumb}  to\textit{Dandy Walker malformation}. Unfortunately we didn't find this evidences in literature.

About Rule (7Z) (HP:0000269, HP:0001305) and the Rule (10Z) (HP:0000567, HP:0001305) \textit{Prominent occiput} (HP:0010636 term) and the  \textit{Chorioretinal colo- ~ boma} (HP:0000567) are the abnormalities related to the \textit{Dandy Walker malformation} as reported in \cite{archer1978enlarged} and \cite{dobyns1989diagnostic}.

\section{Conclusion}
\label{sec:conclusion}
We presented a new methodology based on weighted association rule for HPO data analysis that takes into account the relevance of terms;  the relevance is a weight assigned to a term based on, for example, its specificity to describe a phenotypic abnormality. The relevance of a HPO term, is obtained by computing the IC value related with each term. We presented the outline of an algorithm called HPO-Miner to mine weighted itemsets that have sufficient weighted supports. These itemsets are used in turn to generate association rules that have high weighted support. Finally, the relevance of the mined rules by HPO-Miner, is proved by the evidences found analyzing the literature.


\newpage

\flushbottom

\end{document}